\documentclass{aa}

\usepackage{psfig,subfigure,mncite}

\newcommand{\bvec}[1]{\mbox{\boldmath ${#1}$}}

\begin{document}
    
\title{Coronal stripping in supersaturated stars}

\author{M. Jardine}

\institute{
School of Physics and Astronomy,
North Haugh, St. Andrews KY16 9SS, Scotland
}

\date{Received 2003; accepted 2003}

 
\abstract{
A recent unambiguous detection of X-ray rotational
modulation of the supersaturated star VXR45 (P = 0.223 days) has shown
that its corona has discrete dark and bright X-ray regions.  We suggest
that due to the rapid rotation, the X-ray emitting corona has
been centrifugally stripped away, creating open field regions that are
dark in X-rays.  This leads naturally both to a significant rotational
modulation in X-rays but also to the lower X-ray luminosity of
supersaturated stars compared to those rotating more slowly.  To
demonstrate the effect, we take as an example a more slowly rotating
star for which surface magnetograms are available.  We extrapolate the
potential coronal magnetic field based on these magnetograms and
determine for a hydrostatic, isothermal atmosphere the structure of
the density and of the optically-thin X-ray emission.  We show that if
the rotation rate of this star were increased, the magnitude of the
X-ray luminosity would decrease while its rotational modulation would
increase in a way that is consistent with the recent observations of
VXR45.

}

\maketitle

\section{Introduction}

While the nature of the X-ray emission from the Sun is relatively well
understood, the way in which that emission changes with increasing
rotation rate is less clear.  Observations suggest that, compared to
the Sun, stars of increasing rotation rate show a rise in their X-ray
emission that reaches a maximum of about $L_\mathrm{x}/ L_\mathrm
{bol}=10^{-3}$ at rotation rates of about $v \sin i = 15-20$kms$^{-1}$
\cite{vilhu84}.  Beyond this rotation rate is the ``saturated'' regime
where the X-ray luminosity is independent of rotation rate.  This
behaviour persists until rotation rates of about $v \sin i >
100$kms$^{-1}$, where the X-ray luminosity begins to decrease again. 
This regime is referred to as ``supersaturated''
\cite{prosser96,randich98}.

There are several possible explanations for saturation of the X-ray
emission.  It may be that with increasing rotation rate, the dynamo
process itself saturates due to the back-reaction of the field on the
plasma.  Alternatively the increasing dynamo activity may lead to a
complete coverage of the stellar surface in active regions such that
no further increase in X-ray emission is possible \cite{vilhu84}.  A
rise in coronal temperatures above the detection limit might also produce
an apparent saturation.

A detection of rotational modulation of X-ray emission in a saturated
star could potentially address some of these questions.  A lack of any
modulation could be consistent with a corona that is densely packed
with X-ray emitting loops, while the presence of modulation might
provide some clues as to which magnetic structures are dominating the
emission.  Rotation modulation has, however, proved very difficult to
detect, given the high levels of X-ray variability in active stars.

There have been some indications from observations of both binary and
single stars that rotational modulation may be greater for
lower-temperature plasma which has a smaller pressure scale height and
so may be confined closer to the stellar surface
\cite{white90,gudel95}.  Lack of rotational modulation may of course
be due to an extended, densely-packed corona, but it can also be due
to most of the emission coming from high latitudes where it is not
eclipsed \cite{siarkowski96,solanki97,jeffries98,stepien01}.  A very
clear example of this is the case of AB Dor (P=0.514 days).  A long
term ROSAT study \cite{kurster97} showed a very low rotational
modulation of only $5-13\%$, while observations of two flares with
BeppoSAX \cite{maggio2000} showed no rotational modulation of the
X-ray emission over the decay phase of the flares, although they
lasted for more than one rotation period.  Modelling of the flare
decay indicated that the flaring loops were small, with a maximum
height of only $0.3$R$_{\star}$.  This implies that the flaring
regions must have been located at latitudes above 60$^{\circ}$ (the
stellar inclination) where they were never eclipsed.

Zeeman-Doppler images of AB Dor \cite{donati97abdor95,donati99abdor96}
certainly show flux at the kilogauss level at these latitudes.  By
extrapolating the coronal magnetic field of these magnetograms and
determining the corresponding X-ray emission from an isothermal,
hydrostatic atmosphere, \scite{jardine02xray} showed that the surface
magnetic maps would lead naturally to most of the emission coming from
the high latitude regions.  This gave a low rotational modulation
coupled with the observed high density and emission measure.

In the case of the supersaturated star VXR45, however, the X-ray
emission is very different.  This star is very rapidly rotating member
of IC 2391 with a spectral type of dG9 and photometric period of 0.223
days \cite{patten96} and a $v\sin i > 200$kms$^{-1}$ \cite{stauffer97}. 
Its value of log($L_\mathrm{x}/L_\mathrm{bol}$) is between $-3.60$ 
and $-3.62$ and so it
is clearly in the supersaturated regime.  The XMM-Newton observations
reported by \scite{marino03VXR45} show an unambiguous rotational
modulation of about $30\% $.  Given the broad energy bandpass of
XMM-Newton, \scite{marino03VXR45} rule out the possibility that the
temperature range of the emission is responsible for the
supersaturation and claim that the result shows clearly that the X-ray
emission is coming from discrete structures that are rotationally
self-eclipsed.

It is possible that for these very rapid rotators the dynamo process
has changed its nature from a saturated state to one where it produces
a more patchy coverage of flux at the stellar surface.  Doppler images
of VXR45 do not appear to support this suggestion \cite{marsden2003},
but even if this were the case, the effect of coronal stripping
\cite{jardine99stripping} could mask the onset of such a change, or
disguise it completely.  At high rotation rates the co-rotation radius
(where centrifugal forces balance gravity) moves inside the X-ray
emitting corona.  The rise of gas pressure in the summits of magnetic
loops then breaks open these loops to form open field regions that are
dark in X-rays.  This reduction in the emitting volume initially
balances the rise in the density of the corona to give a saturation of
the X-ray emission, but eventually enough of the coronal volume has
been forced open that the X-ray emission falls with rotation rate.  As
a result, at the highest rotation rates much of the corona is filled
with open field and so there should be a significant rotational
modulation in X-rays.

The purpose of this paper is to demonstrate that a simple increase in
the stellar rotation rate is sufficient to produce both a rise in
rotational modulation and a drop in the emission measure consistent
with that seen in both AB Dor and VXR45.

\section{Modelling the coronal emission}

\begin{figure}

\psfig{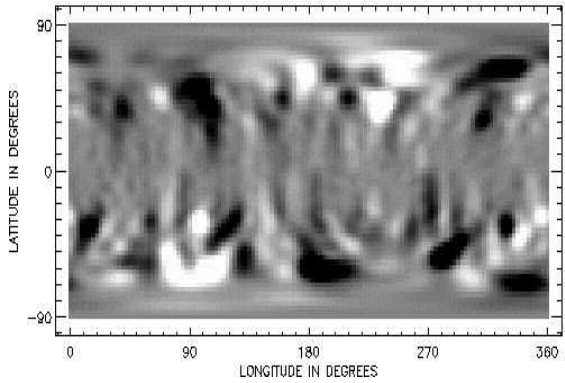}

 \caption{A map of the surface radial magnetic field of AB Dor.  White
 represents $-800$G and black represents 800G. Since AB Dor is inclined
 at 60$^{\circ}$ to the observer, Zeeman-Doppler images provide only
 limited information in the lower hemisphere.  In order to compensate
 for this, we have generated this combined surface map, with the 1995
 map in the upper hemisphere and the 1996 map in the lower hemisphere. 
  }

  \label{surface}
 
\end{figure}
\begin{figure*}
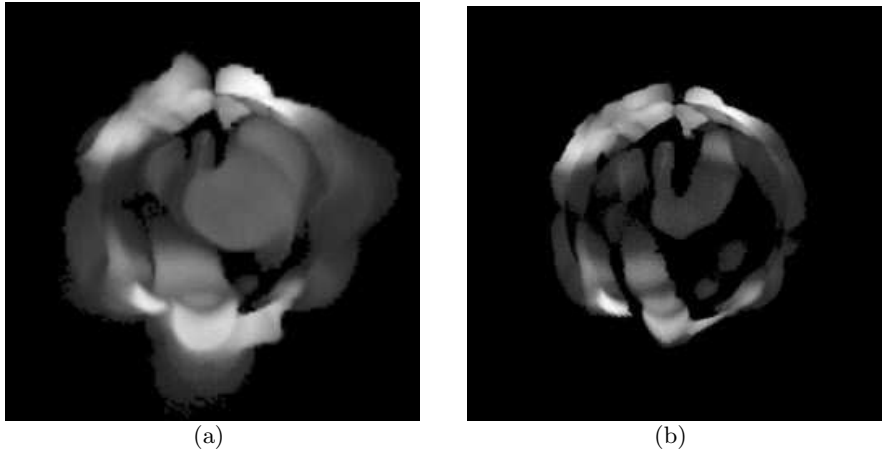

	\def\subfigtopskip{4pt}
	\def\subfigbottomskip{4pt}
	\def\subfigcapskip{2pt}
	\centering
	\begin{tabular}{cc}
		\subfigure[]{
			\label{image_slow} 						
			\psfig{figure=Fi022_f2a.epsf,width=5.5cm}
			} &     	        
		\subfigure[]{
			\label{image_fast} 			
			\psfig{figure=Fi022_f2b.epsf,width=5.5cm}
			} \\
	\end{tabular} 
	
\caption[]{Emission measure images at temperatures of 10$^{7}$K based
on the surface magnetogram shown in Fig.  \ref{surface}.  Two rotation
periods are shown: 0.54 days (left) and 0.17 days (right).  }

    \label{images}
\end{figure*}

We do not have a map of the surface magnetic field of VXR45, although
Doppler images \cite{marsden2003} show a distribution of dark spots
essentially similar to that found on the more slowly rotating star AB
Dor, with much of the surface, especially the high latitude regions,
showing large spots.  Given that we do not know in detail how the
surface flux distribution should vary with rotation rate for these
saturated and supersaturated stars, we choose to take one sample surface
flux distribution and keep that independent of rotation rate.  To this
end we have used a magnetogram for AB Dor to provide a coronal field
structure with the degree of complexity that is implied by
Zeeman-Doppler maps \cite{donati97abdor95,donati99abdor96}.  We note
that due to the stellar inclination of AB Dor, only one hemisphere can
be observed and so to allow for the flux from the hidden hemisphere we
artificially add in a surface map from another year (see Fig
\ref{surface}).  Using this composite surface map as a boundary condition for
the magnetic field, we can extrapolate the coronal field by assuming
it to be potential.

The method of extrapolating the coronal field has been described in
\scite{jardine01structure} and will not be repeated in detail here. 
We use the source surface method pioneered by \scite{altschuler69} and
a code originally developed by \scite{vanballegooijen98}.  Briefly, we
write the magnetic field $\bvec{B}$ in terms of a flux function $\Psi$
such that $\bvec{B} = -\bvec{\nabla} \Psi$ and the condition that the
field is potential ($\bvec{\nabla}\times\bvec{B} =0$) is satisfied
automatically.  The condition that the field is divergence-free then
reduces to Laplace's equation $\bvec{\nabla}^2 \Psi=0$ with solution 
in spherical co-ordinates $(r,\theta,\phi)$
\begin{equation}
 \Psi = \sum_{l=1}^{N}\sum_{m=-l}^{l} [a_{lm}r^l + b_{lm}r^{-(l+1)}]
         P_{lm}(\theta) e^{i m \phi},
\end{equation}
where the associated Legendre functions are denoted by $P_{lm}$.  The
coefficients $a_{lm}$ and $b_{lm}$ are determined by imposing the
radial field at the surface from the Zeeman-Doppler maps and by
assuming that at some height $R_\mathrm{s}$ above the surface the
field becomes radial and hence $B_\theta (R_\mathrm{s}) = 0$.  This
second condition models the effect of the plasma pressure in the
corona pulling open field lines to form a stellar wind.  We set the
source surface to be at the Keplerian co-rotation radius such that for
a stellar rotation rate $\omega$,
$R_\mathrm{s}=(GM/\omega^{2})^{1/3}$.  Thus,
$R_\mathrm{s}=2.71R_{\star}$ for AB Dor and $1.55R_{\star}$ for VXR45.

\begin{figure}

\psfig{figure=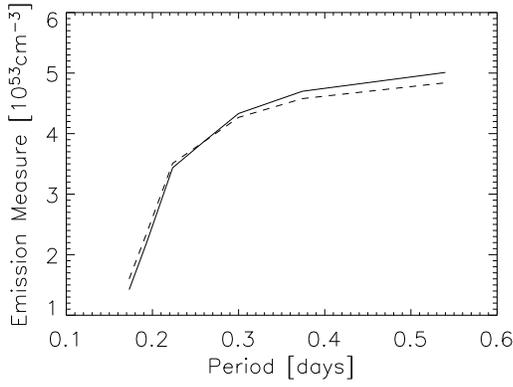,width=8cm}

 \caption{Shown is the emission measure based on the magnetogram shown
 in Fig \ref{surface}.  Results are shown for two assumed stellar
 inclinations, $90^{\circ}$ (solid) and $60^{\circ}$ (dashed).  }

  \label{em}
 
\end{figure}
\begin{figure}

\psfig{figure=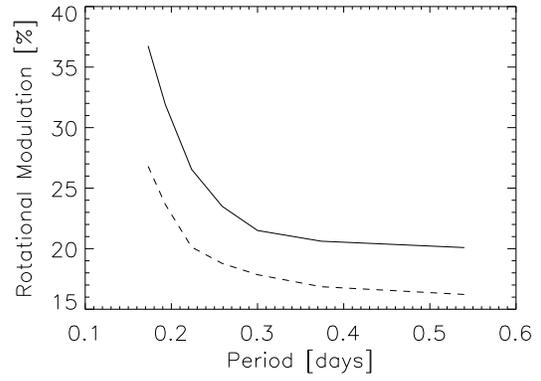,width=8cm}

 \caption{Shown is the rotational modulation of the X-ray emission 
 measure based on the magnetogram shown in Fig \ref{surface}. Results 
 are shown for two assumed stellar inclinations, $90^{\circ}$ (solid) 
 and $60^{\circ}$ (dashed).
  }

  \label{rotmod}
 
\end{figure}

In order to calculate the X-ray emission, we must determine the
coronal density structure.  As a first step, we calculate the pressure
structure of the corona assuming it to be isothermal and in
hydrostatic equilibrium.  Hence the pressure at any point is
$p=p_{0}e^{m/kT\int g_\mathrm{s}ds}$ where $g_\mathrm{s} =( {\bf
g.B})/|{\bf B}|$ is the component of gravity (allowing for rotation)
along the field and
\begin{equation}
g(r,\theta) = \left( -GM_{\star}/r^{2} + 
                     \omega^{2}r\sin^{2} \theta,
		     \omega^{2}r\sin \theta \cos\theta 
             \right).    
\end{equation}
At the loop footpoints we scale the plasma pressure $p_{0}$ to the
magnetic pressure such that $p_{0}(\theta,\phi)=K
B^{2}_{0}(\theta,\phi)$.  We set $K=10^{-5}$ for all models in this
paper, giving a pressure of 120 dyne cm$^{-3}$ for AB Dor at $10^{7}$K
based on an emission-measure weighted density of $4.4\times
10^{10}$cm$^{-3}$.  As described in \scite{jardine02xray} scaling K
simply scales the mean density and the emission measure.  The plasma
pressure within any volume element of the corona is set to zero if the
field line through that volume element is open, or if the gas pressure
exceeds the magnetic pressure (i.e. $\beta>1$) at any point along the
field line.  From the pressure, we calculate the density assuming an
ideal gas and determine the morphology of the optically thin X-ray
emission by integrating along lines of sight through the corona.  We
show results for two stellar inclinations, $60^{\circ}$ (appropriate
for AB Dor) and the value of $90^{\circ}$ determined by
\scite{marino03VXR45} for VXR45.  The rotational modulation is defined
as the fractional change in emission measure over one rotation.

\section{Results and conclusions}

We show in Fig.  \ref{images} images of the X-ray emission for two
sample rotation periods, 0.514 and 0.17 days.  As the scale height
$\Lambda$ has increased with rotation rate from $0.8R_{\star}$ to
$1.3R_{\star}$, more field lines have been forced open by the pressure
of the coronal gas and there is a clear reduction in the volume of the
corona that is bright.  To quantify this, we define a filling factor
$f$ based on the emission-measure-weighted density $\bar{n}_\mathrm{e}=\int
n_\mathrm{e}^{3} dV/\int n_\mathrm{e}^{2}dV$ such that
\begin{equation}
f=\frac{\int n_\mathrm{e}^{2} dV}
  {\frac{4}{3}(R_\mathrm{s}^{3}-R_{\star}^{3})\bar{n}_\mathrm{e}^{2}}.
\end{equation}
If the rotation period decreases from 0.514 to 0.17 days, $f$ falls from
$4.7\times 10^{-2}$ to $1.8\times 10^{-2}$ while the heights of the
brightest regions fall from about $\Lambda/2$ to $\Lambda/10$.

As shown in Figs.  \ref{em} and \ref{rotmod} there is a corresponding
drop in the magnitude of the emission measure but a rise in its
rotational modulation.  The variation of the emission measure for
different scalings of the magnetic field strength and the temperature
with rotation rate has already been explored in
\scite{jardine99stripping}.  The overall level of emission depends on
$n_\mathrm{e}^{2}$ where $n_\mathrm{e} \propto p/T \propto B^{2}/T$, but the
rotational modulation depends on the pattern of bright and dark
regions.  Even if the field strength and temperature are independent
of rotation rate (as in Figs. \ref{em} and \ref{rotmod}) the rise in
the pressure scale height with rotation rate will force closed loops
to become open when $2\mu p > B^{2}$ and so convert bright regions to
dark regions.  Changing the way that the field strength scales with
rotation rate alters the overall level of emission, but has little
effect on the rotational modulation since we have already set
$p\propto B^{2}$ at the coronal base and so the gas and magnetic
pressures rise together.  Lowering the temperature but keeping it and
the field strength independent of rotation rate simply shifts the
curves in Figs.  \ref{em} and \ref{rotmod} to higher values.

The rotational modulation is less for the lower inclination case since
the brightest regions are at high latitude where they remain in view
as the star rotates.  In each hemisphere the brightest region is a
high-latitude section of the large magnetic arcade that runs
north-south over the rotation pole.  Consequently, with an inclination
of 90$^{\circ}$ these two bright regions give two maxima in the
variation of the emission measure with rotational phase (unless they
have been positioned at the same longitude when the composite surface
map is made in which case only one peak is seen).  For lower
inclinations, the bright region in the lower hemisphere is hidden, and
only a single peak remains.

For the case of a $90^{\circ}$ inclination and rotation period of
0.223 days appropriate for VXR45, the rotational modulation is 26$\%$
at $10^{7}$K and 36$\%$ at $10^{6}$K, both close to the observed value
of $\approx 30\%$.  The similarity between the two temperatures (also
noted by \scite{marino03VXR45}) is because the same structures are
emitting at both temperatures.  Since the hotter gas has a greater
scale height, however, it suffers more from coronal stripping and so
the emission measure at $10^{7}$K is about $3\%$ of its value at
$10^{6}$K.

This model is of course not intended to reproduce exactly the observed
results for VXR45.  Its aim is simply to demonstrate the principle
that rapid rotation alone can account for the high rotational
modulation and reduced emission measure observed for a very
rapidly-rotating star.  At high rotation rates the increased plasma 
pressure in the summits of the tallest magnetic loops forces these 
loops to be open, with the result that X-ray bright loops become X-ray 
dark open field regions. Thus, at the highest rotation rates, much of 
the coronal volume is dark in X-rays leading to an increased 
rotational modulation.  


\section{Acknowledgements}
We would like to thank Drs A. van Ballegooijen and K. Wood for allowing
us to use their codes for calculating the potential field extrapolation 
and the optically-thin emission measure respectively.





\end{document}